\documentclass[twocolumn,showpacs,preprintnumbers,amsmath,amssymb]{revtex4}

\usepackage{graphicx}
\usepackage{bm}
\usepackage{color}
\usepackage{amsmath}
\usepackage{amsfonts}
\usepackage{amssymb}
\usepackage{epstopdf}
\usepackage{dcolumn}
\usepackage{here}

\begin{document}

\preprint{APS/123-QED}

\title{Magnetic Structure of the $S\,{=}\,1/2$ Quasi-Two-Dimensional Square-Lattice Heisenberg\\ Antiferromagnet Sr$_2$CuTeO$_6$}

\author{Tomoyuki Koga$^1$, Nobuyuki Kurita$^1$, Maxim Avdeev$^2$, Sergey Danilkin$^2$, Taku J. Sato$^3$, and Hidekazu Tanaka$^1$\thanks{E-mail address: tanaka@lee.phys.titech.ac.jp}}
\affiliation{
$^1$Department of Physics, Tokyo Institute of Technology, Meguro-ku, Tokyo 152-8551, Japan\\
$^2$Bragg Institute, Australian Nuclear Science and Technology Organisation, Lucas Heights, NSW 2234, Australia \\
$^3$Institute of Multidisciplinary Research for Advanced Materials, Tohoku University, Sendai, Miyagi 980-8577, Japan
}
\date{\today}

\begin{abstract}

The magnetic structure of the double perovskite compound Sr$_2$CuTeO$_6$ was determined from neutron powder diffraction data. This material is magnetically described as an $S\,{=}\,1/2$ quasi-two-dimensional square-lattice Heisenberg model with antiferromagnetic nearest-neighbor and next-nearest-neighbor interactions. Sr$_2$CuTeO$_6$ undergoes a magnetic phase transition at $T_{\rm N}\,{\simeq}\,29\,$ K. The spin structure below $T_{\rm N}$ is N\'{e}el antiferromagnetic on the square lattice, which means that the nearest-neighbor interaction ($J_1$) is stronger than the next-nearest-neighbor interaction ($J_2$), in contrast to other isostructural compounds such as Ba$_2$CuWO$_6$ and Sr$_2$CuWO$_6$, for which $|J_1|\,{<}\,|J_2|$ is realized.

\end{abstract}

\pacs{75.10.Jm, 75.25.-j, 75.47.Lx, 75.50.Ee}

\maketitle

\section{Introduction}
One of the most interesting topics in condensed matter physics is the appearance of quantum disordered ground states in low-dimensional frustrated spin systems. An $S\,{=}\,1/2$ square-lattice Heisenberg antiferromagnet (SLHAF) with the nearest-neighbor (NN) interaction $J_1$ and next-nearest-neighbor (NNN) interaction $J_2$ is a typical frustrated quantum magnet. This system, referred to as the $S\,{=}\,1/2$ $J_1\,{-}\,J_2$ SLHAF model, has been theoretically predicted to exhibit a quantum disordered ground state in the range of $\alpha_{c1}\,{<}\,J_2/J_1\,{<}\,\alpha_{c2}$ with $\alpha_{c1}\,{\simeq}\,0.4$ and $\alpha_{c2}\,{\simeq}\,0.6$ \cite{Chandra,Dagotto,Figueirido,Read,Igarashi,Einarsson,Zhitomirsky,Bishop,Sirker,Mambrini,Darradi}. For $J_2/J_1\,{<}\,\alpha_{c1}$ and $\alpha_{c2}\,{<}\,J_2/J_1$, the ground states are known to be N\'{e}el antiferromagnetic and collinear antiferromagnetic, respectively. On the experimental side, several materials such as Li$_2$VO$X$O$_4$ ($X\,{=}\,$Si, Ge) \cite{Melzi1,Melzi2,Rosner} and  $A'A''$VO(PO$_4$)$_2$, where $A'$ and $A''$ are Ba and Cd or both Pb, respectively \cite{Nath1,Nath2}, have been investigated from the viewpoint of whether they are $S\,{=}$\,1/2 $J_1\,{-}\,J_2$ SLHAFs. However, the values of $J_2/J_1$ for these materials are out of the critical range.

\begin{figure}[b]
\begin{center}
\includegraphics[width=0.87\linewidth]{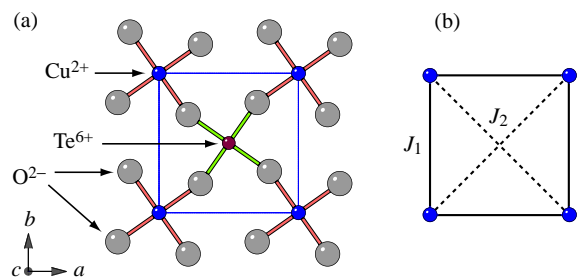}
\end{center}
\vspace{-15pt}\caption{(Color online) (a) Schematic crystal structure of Sr$_2$CuTeO$_6$ in the $c$-plane. (b) Exchange interactions $J_1$ and $J_2$ in the $c$-plane.}
\label{structure}
\end{figure}

The $B$-site ordered double perovskite cuprates $A_2$Cu$M$O$_6$, where $A\,{=}\,$Sr or Ba and $M\,{=}\,$Mo, Te, or W, are also considered to be $S\,{=}\,1/2$ $J_1\,{-}\,J_2$ SLHAFs \cite{Iwanaga}. All these compounds crystallize in the tetragonal structure with space group $I4/m$, in which CuO$_6$ and $M$O$_6$ octahedra are arranged alternately, sharing their corners, as shown in Fig.~\ref{structure}(a). Because these CuO$_6$ octahedra are elongated along the $c$-axis owing to the Jahn-Teller effect, the hole orbitals $d_{x^2-y^2}$ of Cu$^{2+}$ ions with spin-1/2 are spread in the $c$-plane, resulting in  strong intraplane and relatively weak interplane exchange interactions. In the $c$-plane, magnetic Cu$^{2+}$ ions form a uniform square lattice with the NN and NNN exchange interactions via $M$O$_6$ octahedra centered by hexavalent $M^{6+}$ ions (Fig.~\ref{structure}(b)). The low-dimensionality of the system can be inferred from the magnetic susceptibility, which shows a broad maximum at approximately 70\,--\,200\,K \cite{Iwanaga,Todate1,Vasala1}. 

Although such double perovskites are magnetically quasi-two-dimensional (2D) systems, 3D ordering often takes place at low temperatures owing to weak interactions between layers. There are several types of antiferromagnetic long-range order on the square lattice for the ordered double perovskite~\cite{Battle}, depending on the relative strength of interactions ($J_2/J_1$). For $J_2/J_1\,{<}\,1$, the ordering on the square lattice is N\'{e}el antiferromagnetic (NAF), while collinear antiferromagnetic (CAF) ordering takes place for $J_2/J_1\,{>}\,1$. The spin ordering along the $c$-direction depends on the interlayer exchange interactions.

For Ba$_2$CuWO$_6$ and Sr$_2$CuWO$_6$, the magnetic structure has been determined by neutron powder diffraction measurements \cite{Todate2,Vasala2}. The observed magnetic structure of Ba$_2$CuWO$_6$ is described by the propagation vector ${\bm k}\,{=}\,(1/2,\,1/2,\,1/2)$ on the face-centered lattice~\cite{Todate2}. This indicates the CAF order on the square lattice. Recently, Vasala {\it et al.} performed neutron powder diffraction experiments on Sr$_2$CuWO$_6$ and found a magnetic Bragg peak at ${\bm k}\,{=}\,(0,\,1/2,\,1/2)$ on the tetragonal body-centered lattice, indicative of the CAF order~\cite{Vasala2}. Therefore, the spin arrangement on the square lattice is the same as that of Ba$_2$CuWO$_6$. These results revealed that $J_2$ is dominant over $J_1$ in both compounds. For Sr$_2$CuMoO$_6$, although the magnetic structure has not yet been determined, Vasala {\it et al.} predicted that $J_2$ is stronger than $J_1$ from {\it ab initio} calculations combined with X-ray absorption experiments~\cite{Vasala3}. The transition temperatures of Ba$_2$CuWO$_6$, Sr$_2$CuWO$_6$, and Sr$_2$CuMoO$_6$ have been determined from $\mu$SR experiments to be $T_{\rm N}\,{=}\,28$, 24, and 28\,K, respectively~\cite{Todate2, Vasala3}.

In this paper, we report the results of neutron powder diffraction measurements on Sr$_2$CuTeO$_6$. 
It was found that Sr$_2$CuTeO$_6$ exhibits the NAF order, in contrast to other double perovskite cuprates $A_2$Cu$M$O$_6$ with $M\,{=}\,$Mo and W. Additionally, the transition temperature was determined to be $T_{\rm N}\,{=}\,29$\,K. We discuss the mechanism leading to different spin structures between Sr$_2$CuTeO$_6$ and the other $A_2$Cu$M$O$_6$ systems with emphasis on the electronic state of the filled outermost orbital of the nonmagnetic hexavalent ion $M^{6+}$.

\section{Experimental details}

A powder sample of Sr$_2$CuTeO$_6$ was synthesized from a stoichiometric mixture of SrO, CuO, and TeO$_2$ by a solid-state reaction. The mixed powder was ground well with an agate mortar and fired at 900\,$^{\circ}$C in air for 24\,h. The powder was then reground, pelletized, and calcined twice at 1100\,$^{\circ}$C in air for 24\,h.

The neutron powder diffraction experiments were performed using the high-resolution powder diffractometer Echidna installed at the OPAL reactor, ANSTO. The data were collected at 1.5 K and from 10 to 40 K at intervals of 5 K with a neutron wavelength of 2.4395\,\AA. The sample was placed in a cylindrical vanadium can with an aluminum cap. To evaluate the transition temperature more accurately, we measured the temperature dependence of the magnetic peak using the triple-axis spectrometer Taipan. In the diffraction mode of Taipan, a neutron beam with a high flux is available, which is useful for detecting weak magnetic Bragg peaks. Incident neutrons with a wavelength of ${\lambda}\,{=}\,2.345$\,{\AA}  were selected using pyrolytic graphite (PG) 002 monochromator, without any additional collimation, i.e. the ``open\,-\,open\,-\,open\,-\,open'' configuration with the 10\,mm wide neutron beam was employed. To eliminate higher order neutrons, PG filters were placed before and after the sample position. 
The powder diffraction data were analyzed by the Rietveld method using Fullprof software. The magnetic form factor of the Cu$^{2+}$ ion was taken from the literature~\cite{Juan}.

\begin{figure}[t]
\begin{center}
\includegraphics[width=\linewidth]{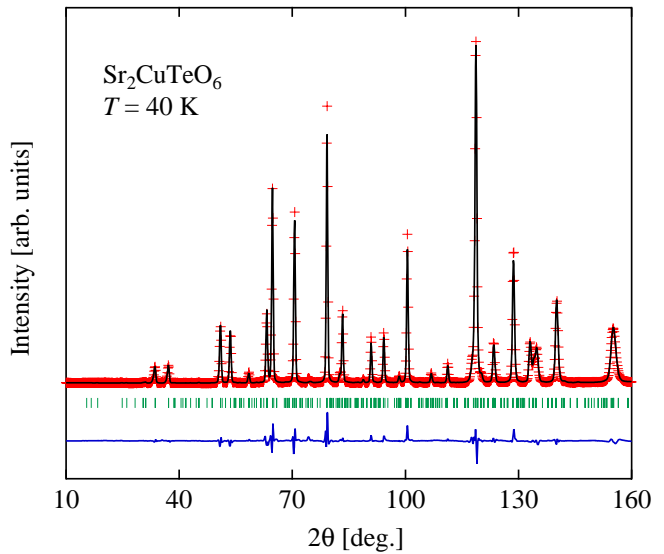}
\end{center}
\vspace{-15pt}\caption{(Color online) Neutron powder diffraction pattern of Sr$_2$CuTeO$_6$ measured at 40\,K (red crosses) and the result of Rietveld fitting (black line). The blue curve shows the difference between them.} 
\label{diffraction}
\end{figure}

\begin{figure}[t]
\begin{center}
\includegraphics[width=\linewidth]{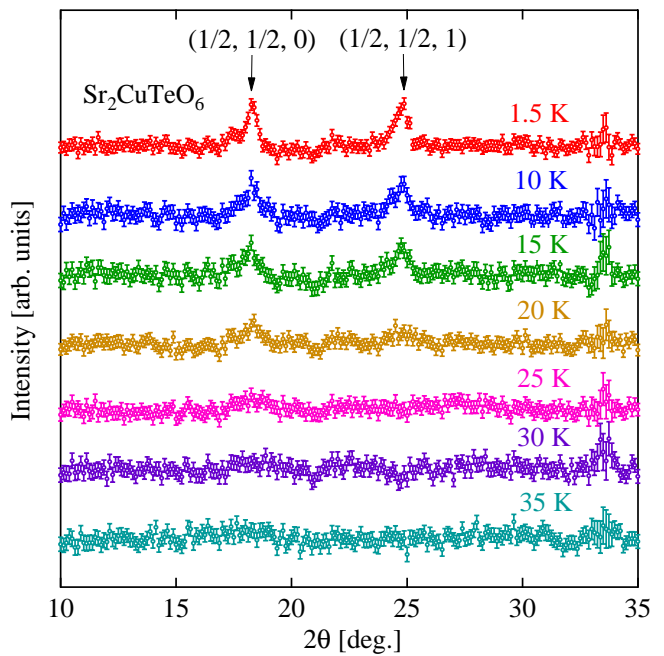}
\end{center}
\vspace{-15pt}\caption{(Color online) Neutron powder diffraction spectra collected at various temperatures, where the diffraction spectrum at 40\,K was subtracted as the background. Two magnetic peaks assigned to ${\bm Q}\,{=}\,(1/2,\,1/2,\,0)$ and (1/2,\,1/2,\,1) are observed at $2{\theta}\,{\simeq}\,18$$^{\circ}$ and 25$^{\circ}$, respectively.} 
\label{magnetic_peaks}
\end{figure}

\begin{table*}[t]
   \renewcommand{\arraystretch}{1.2}
   \caption{Refined structural and magnetic parameters for Sr$_2$CuTeO$_6$ with space group $I4/m$ based on neutron powder diffraction data.}\vspace{2mm} \label{parameter}
\begin{tabular}{lD{.}{.}{5}D{.}{.}{5}D{.}{.}{5}D{.}{.}{5}D{.}{.}{5}D{.}{.}{5}D{.}{.}{5}D{.}{.}{5}}\hline \hline
\multicolumn{1}{l}{$T$\,(K)} &  \multicolumn{1}{c}{\hspace{20pt}1.5} & \multicolumn{1}{c}{\hspace{20pt}10} & \multicolumn{1}{c}{\hspace{20pt}15} & \multicolumn{1}{c}{\hspace{20pt}20} & \multicolumn{1}{c}{\hspace{20pt}25} & \multicolumn{1}{c}{\hspace{20pt}30} & \multicolumn{1}{c}{\hspace{20pt}35} & \multicolumn{1}{c}{\hspace{20pt}40} \\ \hline
$a$\,(\AA) & \hspace{20pt}5.4184(1) & \hspace{20pt}5.4185(1) & \hspace{20pt}5.4184(2) & \hspace{20pt}5.4185(1) & \hspace{20pt}5.4185(1) & \hspace{20pt}5.4185(2) & \hspace{20pt}5.4186(2) & \hspace{20pt}5.4186(1) \\
$c$\,(\AA) & 8.4521(2) & 8.4522(2) & 8.4521(2) & 8.4522(2) & 8.4522(2) & 8.4522(2) & 8.4524(2) & 8.4524(2) \\
$V$\,(\AA$^3$) & 248.15(1) & 248.16(1) & 248.15(1) & 248.16(1) & 248.16(1) & 248.16(2) & 248.17(3) & 248.17(2) \\
Sr(0, 0.5, 0.25)\\
$B$\,(\AA$^2$) & 0.07(6) & 0.09(6) & 0.07(6) & 0.08(6) & 0.07(5) & 0.08(6) & 0.09(6) & 0.09(6) \\
Cu(0, 0, 0)\\
$B$\,(\AA$^2$) & 0.05(7) & 0.07(8) & 0.04(8) & 0.06(7) & 0.05(7) & 0.06(8) & 0.03(8) & 0.06(8) \\
Te(0, 0, 0.5)\\
$B$\,(\AA$^2$) & 0.03(10) & 0.04(10) & 0.05(10) & 0.05(10) & 0.03(10) & 0.04(10) & 0.07(10) & 0.03(10) \\
O(1)($x$, $y$, 0)\\
$x$ & 0.3026(4) & 0.3027(5) & 0.3028(5) & 0.3027(4) & 0.3028(4) & 0.3027(5) & 0.3028(5) & 0.3028(5) \\
$y$ & 0.2073(4) & 0.2073(4) & 0.2073(4) & 0.2073(4) & 0.2074(4) & 0.2073(4) & 0.2075(4) & 0.2073(4) \\
$B$\,(\AA$^2$) & 0.14(5) & 0.16(5) & 0.15(5) & 0.16(5) & 0.15(4) & 0.16(5) & 0.15(5) & 0.16(5) \\
O(2)(0, 0, $z$)\\
$z$ & 0.2750(2) & 0.2750(2) & 0.2751(2) & 0.2751(2) & 0.2750(2) & 0.2751(2) & 0.2750(2) & 0.2750(2) \\
$B$\,(\AA$^2$) & 0.24(7) & 0.26(7) & 0.24(7) & 0.25(7) & 0.25(7) & 0.25(7) & 0.27(7) & 0.26(7) \\
\\
$m\,(\mu_{\rm B})$ & 0.687(60) & 0.628(65) & 0.588(69) & 0.496(80) & 0.349(116) & - & - & - \\
\\
$R_p$\,(\%) & 7.84 & 8.05 & 8.08 & 8.01 & 7.92 & 8.06 & 8.02 & 8.02 \\
$R_{wp}\,(\%)$ & 11.7 & 11.9 & 12.0 & 11.8 & 11.5 & 11.9 & 11.8 & 11.7 \\
\hline \hline
 \end{tabular}
\end{table*}

\section{Results}
Figure~\ref{diffraction} shows the neutron powder diffraction pattern for Sr$_2$CuTeO$_6$ collected at 40\,K ($\,{>}\,T_{\rm N}\,{=}\,29$\,K) and the result of Rietveld analysis. The analysis was based on the structural model with space group $I4/m$ determined at room temperature by Iwanaga {\it et al.}~\cite{Iwanaga}. We used this model as an initial model because we confirmed that there is no structural transition below room temperature by magnetic susceptibility and specific heat measurements~\cite{Koga}. To achieve a better fit to the experimental data, the small half-lambda contribution ($\sim 0.1$\%) and the contaminating aluminum reflections from the cryostat walls are taken into account. The structural and magnetic parameters refined at various temperatures are summarized in Table~\ref{parameter}. All the structural parameters are almost independent of the temperature below 40\,K.

At low temperatures, two additional peaks, indicative of 3D magnetic long-range order in Sr$_2$CuTeO$_6$, clearly appear in the neutron diffraction patterns at $2{\theta}\,{\simeq}\,18$$^{\circ}$ and 25$^{\circ}$. Figure~\ref{magnetic_peaks} shows neutron powder diffraction spectra measured at various temperatures focusing on these two peaks, where the diffraction spectrum at 40\,K has been subtracted so that only magnetic Bragg peaks are visible. These two magnetic peaks can be assigned to $\bm Q$\,{=}\,(1/2,\,1/2,\,0) and (1/2,\,1/2,\,1) diffraction peaks on the tetragonal unit cell, respectively. Therefore, the magnetic ordering in Sr$_2$CuTeO$_6$ is NAF on the square lattice characterized by the propagation vector $\bm k$\,{=}\,(1/2,\,1/2,\,0).

\begin{figure}
\begin{center}
\includegraphics[width=0.877\linewidth]{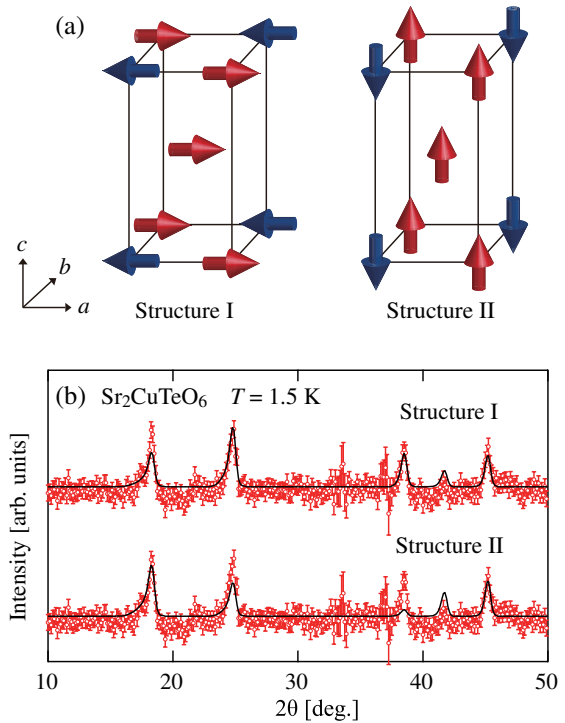}
\end{center}
\vspace{-15pt}\caption{(Color online) (a) Possible magnetic structures I and II for Sr$_2$CuTeO$_6$. (b) Magnetic diffraction pattern measured at 1.5\,K, where the nuclear diffraction pattern at 40\,K was subtracted, and the fits with structures I and II, for which $\chi^2\,{=}\,1.11$ and 1.83, respectively.}
\label{configuration}
\end{figure}

To determine the magnetic structure, we analyzed the magnetic diffraction pattern at 1.5\,K, where the nuclear diffraction pattern at 40\,K was subtracted. There are two possible spin configurations depending on whether all the spins are directed along the $a$-axis (structure I) or the $c$-axis (structure I\hspace{-1pt}I) as shown in Fig.~\ref{configuration}(a). The $b$-axis is equivalent to the $a$-axis owing to the symmetry of the crystal structure and the magnetic propagation vector. We fitted the calculated diffraction patterns for structures I and I\hspace{-1pt}I to the experimental magnetic diffraction pattern using the structural parameters and scale factor determined by Rietveld analysis at 40\,K. The fitting range is limited to vicinities of the expected magnetic Bragg peaks below 2$\theta$\,{=}\,50$^{\circ}$ to eliminate the effects from large errors of nuclear peaks and background. As shown in Fig.~\ref{configuration}(b), structure I reproduces the experimental diffraction pattern better than structure I\hspace{-1pt}I. The values of ${\chi}^2$ are 1.11 and 1.83 for structures I and I\hspace{-1pt}I, respectively. From this result, we conclude that structure I is realized in the ordered state of Sr$_2$CuTeO$_6$. 

\begin{figure}
\begin{center}
\includegraphics[width=\linewidth]{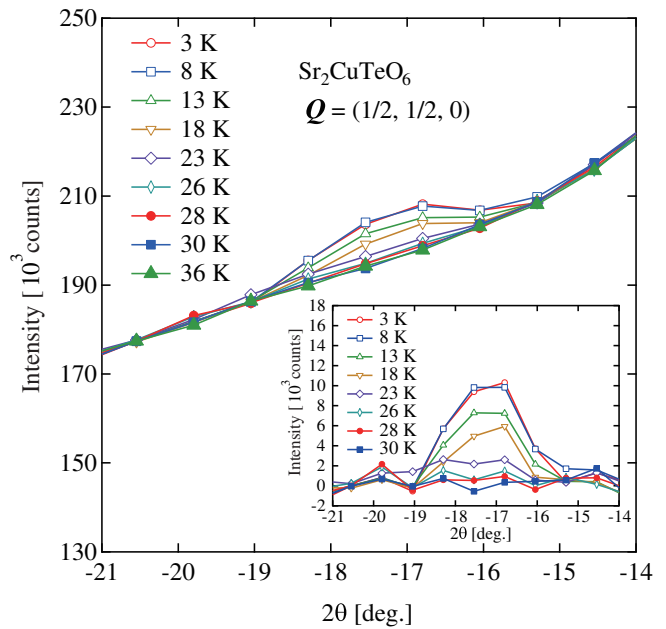}
\end{center}
\vspace{-15pt}\caption{(Color online) Neutron powder diffraction profiles for ${\bm Q}\,{=}\,(1/2,\,1/2,\,0)$ magnetic Bragg peak for Sr$_2$CuTeO$_6$ measured at various temperatures using the triple-axis spectrometer Taipan. The inset shows the magnetic peak intensities obtained by subtracting the data at 36\,K as the background.} \label{triple-axis}
\end{figure}

\begin{figure}
\begin{center}
\includegraphics[width=\linewidth]{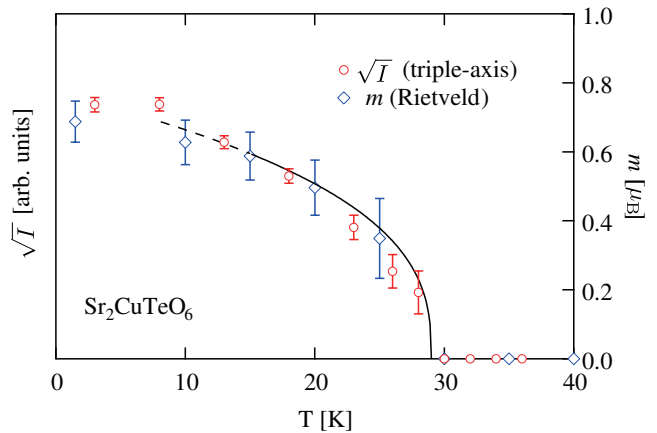}
\end{center}
\vspace{-15pt}\caption{(Color online) Temperature dependence of the normalized square root of the magnetic peak intensity $\sqrt{I}$ for ${\bm Q}\,{=}\,(1/2,\,1/2,\,0)$ diffraction in Sr$_2$CuTeO$_6$.  The blue squares show the ordered moment $m$ obtained from Rietveld analysis (right axis). The solid line is a visual guide.} \label{moment}
\end{figure}

As shown in Fig.~\ref{magnetic_peaks}, the magnetic peaks vanish between 20 and 30\,K. To determine the transition temperature, we also collected diffraction data of Sr$_2$CuTeO$_6$ using the triple-axis spectrometer Taipan in the diffraction mode. Figure~\ref{triple-axis} shows the temperature evolution of the magnetic Bragg peak for ${\bm Q}$\,{=}\,(1/2,\,1/2,\,0). The background is rather high and is sloped because of the tail of the direct beam. The peak height decreases continuously as the temperature increases and becomes undetectable at $T\,{\ge}\,$30\,K. The diffraction data at 36\,K was assumed as the background. The inset of Fig.~\ref{triple-axis} shows the magnetic peak intensities obtained by subtracting the data at 36\,K. The magnetic peak intensity was evaluated by Gaussian fitting, where only height was a free parameter and others were fixed to the results of the fitting for the data at 3\,K. Figure~\ref{moment} shows the temperature dependence of the normalized square root of the magnetic peak intensity $\sqrt{I}$ for ${\bm Q}$\,{=}\,(1/2,\,1/2,\,0) diffraction. Actually, it was difficult to fit the data at $T\,{\ge}\,$30\,K correctly because the peak is smaller than large errors, so that $\sqrt{I}$ was defined to be zero. 
The transition temperature was thus evaluated to be $T_{\rm N}\,{\simeq}\,$29\,K~\cite{notation}. 

The magnitude of the ordered moment $m$ in Sr$_2$CuTeO$_6$ below $T_{\rm N}$ was evaluated from the powder Rietveld analysis using Fullprof software on the basis of spin structure I. Results are shown in Table \ref{parameter} together with the crystal structure parameters. The magnitude of the ordered moment  at 1.5\,K was obtained as $m\,{=}\,0.69(6)$\,$\mu_{\rm B}$, which is larger than $m\,{=}\,0.2$ and 0.57\,$\mu_{\rm B}$ reported for Ba$_2$CuWO$_6$~\cite{Todate2} and Sr$_2$CuWO$_6$~\cite{Vasala2}, respectively. The temperature dependence of the ordered moment $m$ is plotted in Fig.~\ref{moment}. Above 20\,K, the normalized square root of the magnetic peak intensity $\sqrt{I}$ obtained from the triple-axis data is considerably smaller than the ordered moment $m$. 


\section{Discussion}
In this work, we found that the magnetic structure in the ordered state of Sr$_2$CuTeO$_6$ is NAF,  in contrast to the CAF order in other double perovskite cuprates $A_2$Cu$M$O$_6$ with $M\,{=}\,$Mo and W, where $|J_2|\,{>}\,|J_1|$~\cite{Todate2,Vasala2}. This indicates that the condition $J_1\,{>}\,J_2$ is realized in Sr$_2$CuTeO$_6$. This difference can be understood by considering the superexchange interactions via the hexavalent $M^{6+}$ ion, as shown below.


We discuss the superexchange interactions in accordance with Kanamori theory~\cite{Kanamori}.
In the $A_2$Cu$M$O$_6$-type double perovskite compounds, one of the dominant paths of  the NN superexchange interaction $J_1$ is Cu$^{2+}-$\,O$^{2-}-$ O$^{2-}-$\,Cu$^{2+}$, which is common to $A_2$Cu$M$O$_6$ compounds. The superexchange interaction via this path should be antiferromagnetic. The other dominant path is considered to be Cu$^{2+}$${-}$\,O$^{2-}$${-}$\,$M^{6+}$${-}$\,O$^{2-}$${-}$\,Cu$^{2+}$ because for $M$\,=\,W, the dominant NNN interaction $J_2$ acting between spins located on the diagonal lattice points of the square lattice is mediated by the intermediate WO$_6$ octahedron, which shares a corner with the CuO$_6$ octahedron. The sign of the superexchange interaction in principle depends on the filled outermost orbital of the nonmagnetic hexavalent $M^{6+}$ ion, which is the 4$d$ orbital for $M$\,=\,Te and the 4$p$ or 5$p$ orbital for $M$\,=\,Mo and W.

\begin{figure}[t]
\begin{center}
\includegraphics[width=\linewidth]{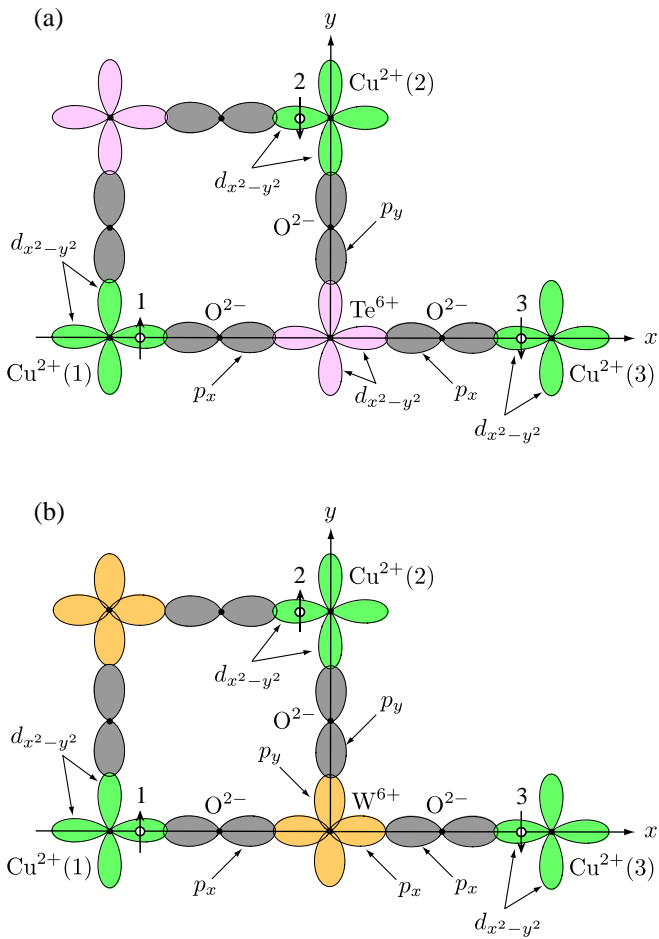}
\end{center}
\vspace{-15pt}\caption{(Color online) Orbital configurations related to superexchange interactions in $A_2$Cu$M$O$_6$ via Cu$^{2+}$$-$\,O$^{2-}$$-$\,M$^{6+}$ $-$\,O$^{2-}$$-$\,Cu$^{2+}$ (a) for $M$\,=\,Te and (b) for $M$\,=\,W.} 
\label{orbital}
\end{figure}

Figure~\ref{orbital} illustrates the orbital configurations related to superexchange interactions for $M$\,=\,Te and $M$\,=\,W. We consider the superexchange interaction between hole spins on the $d_{x^2{-}y^2}$ orbitals of Cu$^{2+}$ ions via $d_{x^2{-}y^2}$ for $M$\,=\,Te and via $p_x$ and $p_y$ for $M$\,=\,W. For simplification, we assume that Cu$^{2+}$$-$\,O$^{2-}$$-$\,$M^{6+}$ is a straight line, although it is actually a zigzag line, as shown in Fig.~\ref{structure}(a). The superexchange interaction $J_1$ between Cu$^{2+}$(1) and Cu$^{2+}$(2) for $M$\,=\,Te is based on the following process.
\begin{enumerate}
\vspace{-2mm}
\renewcommand{\labelenumi}{\arabic{enumi})}
\setlength{\itemsep}{0mm}
\item
The $p_x$ and $p_y$ orbitals of O$^{2-}$ ions overlap strongly with the $d_{x^2{-}y^2}$ orbital of Te$^{6+}$, so that these orbitals together form a molecular orbital. 
\item 
Hole 1 on Cu$^{2+}$(1) and hole 2 on Cu$^{2+}$(2) are transferred to the same molecular orbital. In this case, the two hole spins must be antiparallel owing to the Pauli principle. 
\item 
The two holes are transferred back to the $d_{x^2{-}y^2}$ orbitals of the two Cu$^{2+}$ ions. 
\vspace{-2mm}
\end{enumerate}
This process results in an antiferromagnetic superexchange interaction between Cu$^{2+}$(1) and Cu$^{2+}$(2). Because the same argument is also applicable to the superexchange interaction between Cu$^{2+}$(1) and Cu$^{2+}$(3), the NNN exchange interaction $J_2$ becomes antiferromagnetic. As shown in Fig.~\ref{orbital}, there are two Cu$^{2+}$$-$\,O$^{2-}$$-$ Te$^{6+}$$-$\,O$^{2-}$$-$\,Cu$^{2+}$ paths for $J_1$, whereas for $J_2$, there is a single exchange path. The contributions of the paths to the superexchange interaction should be almost the same. Thus, for the superexchange interaction mediated by the TeO$_6$ octahedron, we can expect $J_1\,{\simeq}\,2J_2$. 

For the superexchange path Cu$^{2+}-$\,O$^{2-}-$\,O$^{2-}-$\,Cu$^{2+}$, the $p_x$ and $p_y$ orbitals of the two O$^{2-}$ ions form a molecular orbital, although their overlap is small. In the molecular orbital, two hole spins transferred from Cu$^{2+}$ ions must be antiparallel owing to the Pauli principle. This makes an antiferromagnetic contribution to the $J_1$ interaction.
Therefore, the relation between $J_1$ and $J_2$ will satisfy $J_1\,{>}\,2J_2$, which is consistent with the NAF state observed in Sr$_2$CuTeO$_6$. 

For $M$\,=\,W and Mo, the superexchange interaction $J_1$ between Cu$^{2+}$(1) and Cu$^{2+}$(2) is based on the following process.
\begin{enumerate}
\vspace{-2mm}
\renewcommand{\labelenumi}{\arabic{enumi})}
\setlength{\itemsep}{0mm}
\item
The $p_x$ and $p_y$ orbitals of O$^{2-}$ ions overlap strongly with the $p_x$ and $p_y$ orbitals of W$^{6+}$ (or Mo$^{6+}$), so that these orbitals form two different molecular orbitals, which are orthogonal to each other. 
\item 
Hole 1 on Cu$^{2+}$(1) and hole 2 on Cu$^{2+}$(2) are transferred to different molecular orbitals. In this case, the two hole spins must be parallel owing to the Hund rule for the hole spins on the $p_x$ and $p_y$ orbitals of W$^{6+}$. 
\item 
The two holes are transferred back to the $d_{x^2{-}y^2}$ orbitals of the two Cu$^{2+}$ ions. 
\vspace{-2mm}
\end{enumerate}
This process results in a ferromagnetic superexchange interaction between Cu$^{2+}$(1) and Cu$^{2+}$(2).

For the NNN exchange interaction $J_2$ for $M\,{=}\,$W and Mo, hole 1 on Cu$^{2+}$(1) and hole 3 on Cu$^{2+}$(3) are transferred to the same molecular orbital. In this case, the two hole spins must be antiparallel owing to the Pauli principle. Thus, the $J_2$ interaction becomes antiferromagnetic. For the $J_1$ interaction, the antiferromagnetic contribution from the path Cu$^{2+}-$\,O$^{2-}-$\,O$^{2-}-$ Cu$^{2+}$ and the ferromagnetic contribution from the path Cu$^{2+}$$-$\,O$^{2-}$$-$\,W$^{6+}$$-$\,O$^{2-}$$-$\,Cu$^{2+}$ mostly cancel out, as expected in Ba$_3$CoNb$_2$O$_9$~\cite{Yokota}. This leads to the condition $|J_2|\,{\gg}\,|J_1|$, which stabilizes the CAF state as observed for $M\,{=}\,$ W~\cite{Todate2,Vasala2}. From these arguments, we can deduce that the difference in the filled outermost orbitals of nonmagnetic hexavalent $M^{6+}$ ions gives rise to the different magnetic orderings in the $A_2$Cu$M$O$_6$-type double perovskite compounds.


\section{Conclusion}

The magnetic structure of  Sr$_2$CuTeO$_6$ was determined from neutron powder diffraction measurements. The structure is of the N\'{e}el antiferromagnetic type on the square lattice with the propagation vector $\bm k$\,{=}\,(1/2,\,1/2,\,0). From Rietveld analysis, the ordered moment lies in the $c$-plane and its magnitude was evaluated to be 0.69\,$\mu_{\rm B}$ at 1.5\,K. The transition temperature was also determined to be $T_{\rm N}\,{\simeq}\,29\,$K, which is comparable to those observed in other $A_2$Cu$M$O$_6$-type double perovskite compounds. The condition $J_1\,{>}\,J_2$ for the magnitudes of the NN and NNN exchange interactions for Sr$_2$CuTeO$_6$ is in sharp contrast to the condition $|J_2|\,{>}\,|J_1|$ for other isostructural compounds including Ba$_2$CuWO$_6$ and Sr$_2$CuWO$_6$. The origin of the difference can be attributed to whether the filled outermost orbital of the nonmagnetic hexavalent ion is a $p$ or $d$ orbital. This work can pave the way for creating doping series Sr$_2$CuTe$_{1-x}$W$_x$O$_6$ that can access the quantum disordered region of $\alpha_{c1}\,{<}\,J_2/J_1\,{<}\,\alpha_{c2}$. 


\section*{ACKNOWLEDGMENTS}

This work was supported by Grants-in-Aid for Scientific Research (A) (Grants No. 23244072 and No. 26247058) and a Grant-in-Aid for Young Scientists (B) (Grant No. 26800181) from Japan Society for the Promotion of Science. TJS was partly supported by General User Program for Neutron Scattering Experiments, ISSP, University Tokyo.


\begin{thebibliography}{25}
\bibitem{notation} In our previous investigation \cite{Koga}, we obtained $T_{\rm N}\,{=}\,4.8$\,K for Sr$_2$CuTeO$_6$ from a heat capacity anomaly. However, it was found afterward that the anomaly occurred as a result of the contribution of accidentally synthesized impurities.
\end{thebibliography}
\end{document}